\documentclass[conference]{IEEEtran}
\IEEEoverridecommandlockouts

\usepackage{cite}
\usepackage{amsmath,amssymb,amsfonts}
\usepackage{algorithmic}
\usepackage{graphicx}
\usepackage{textcomp}
\usepackage{xcolor}
\def\BibTeX{{\rm B\kern-.05em{\sc i\kern-.025em b}\kern-.08em
    T\kern-.1667em\lower.7ex\hbox{E}\kern-.125emX}}
\usepackage[utf8]{inputenc}
\usepackage[T1]{fontenc}
\usepackage{booktabs}
\usepackage{siunitx}
\usepackage{tikz}
\usepackage{algorithm}
\usepackage{subcaption}
\usepackage{soul}
\usepackage{makecell}
\usepackage{verbatim}
\title{Aerial Booster-Cell Enabled Inter-Cell Interference Coordination for 5G NR Networks}
\author{
\IEEEauthorblockN{Md Sharif Hossen, Vijay K. Shah and Ismail Guvenc}
\IEEEauthorblockA{Department of Electrical and Computer Engineering\\
North Carolina State University, Raleigh, NC, USA\\
Email: \{mhossen, vijay.shah, iguvenc\}@ncsu.edu}
}

\begin{document}

\maketitle
\begin{abstract}
Cellular-connected unmanned aerial vehicles (UAVs) operating in 5G New Radio (NR) macro networks experience severe and spatially non-uniform downlink interference. This is primarily caused by the interference from the sidelobes of downtilted base station (BS) antennas serving terrestrial users, which limits the ability of the network to provide uniform and high-quality coverage to aerial users. Supporting aerial users requires boosting the coverage of certain cells or sectors, which can further exacerbate inter-cell interference in dense macro deployments. This motivates the need for inter-cell interference coordination (ICIC) in multi-cell 5G NR networks serving both aerial and terrestrial users. In this work, we propose an ICIC framework that jointly optimizes antenna-domain coordination through BS uptilt angle optimization and time-domain interference coordination (TDIC) through NR-compliant scheduling. The framework is formulated as a multi-cell NR macro deployment problem that maximizes the minimum UAV signal-to-interference ratio (SIR) over a spatial grid of UAV locations while maintaining acceptable performance for ground user equipment (GUEs). The resulting optimization problem is non-convex and is solved using bio-inspired optimization techniques, including particle swarm optimization (PSO) and genetic algorithm (GA). Simulation results demonstrate that coordinated uptilt optimization with the booster-cell architecture significantly improves worst-case UAV SIR and downlink reliability in multi-cell 5G NR networks.
\end{abstract}

\section{Introduction}
Cellular-connected unmanned aerial vehicles (UAVs) are envisioned to support a variety of applications, ranging from public safety and infrastructure inspection to wide-area monitoring \cite{Muruganathan_9696263}. Leveraging existing 5G New Radio (NR) macrocellular infrastructure for UAV connectivity is essential to support these applications and to enhance coverage, mobility, and cost efficiency. However, the current network infrastructures are primarily designed for terrestrial users, where base station (BS) antennas are electrically downtilted to improve coverage for ground users. UAVs are often served through antenna sidelobes, and they experience strong line-of-sight (LoS) interference from the sidelobes and ground-scattered reflections \cite{ITU_UAV_Reliability_2021} \cite{donggu_lee}.
Hence, UAV downlink performance is often interference-limited and highly non-uniform across space. 
These spatial reliability extremes dominate safety-critical UAV command-and-control (C2) links \cite{Nassif2024ElectronicConspicuity}
 \cite{Hosseini8741719} in 5G NR \cite{TS22125}, which has not been explicitly addressed in existing studies \cite{Du9815327}, \cite{Maeng10487029}. Also, UAV corridor coverage with BS antenna uptilt has been studied in \cite{Maeng10487029}, showing that upward antenna steering can improve aerial coverage. While Moin et al. \cite{ITU_UAV_Reliability_2021} introduce and validate the uptilted antenna concept, this work advances it into a network-side solution that explicitly targets worst-case downlink signal-to-interference ratio (SIR) through BS antenna uptilt optimization and NR-compatible time-domain coordination, which makes it suitable for safety-critical UAV operations.

Motivated by this observation, we focus on improving the worst-case UAV downlink SIR in multi-cell 5G NR macro networks using network-side control mechanisms. Specifically, we consider a deployable BS-antenna architecture in which each BS is equipped with co-located downtilted and uptilted antenna sectors. We refer to the uptilted antenna sector, operating as a controllable aerial serving entity, as a booster cell. Similar concepts have been explored in cellular networks where additional cells or sectors are deployed to serve localized hotspot traffic and enhance network capacity through network densification \cite{6736747_network_densification}, \cite{6171992_femtocells}. The booster cell is not a separate base station or carrier but a co-located antenna system that provides enhanced aerial coverage without affecting traditional terrestrial operations. Hence, the downtilted sector maintains legacy terrestrial coverage, while the uptilted sector provides controlled energy pointing towards aerial users without modifying the NR air interface or UAV hardware. This dual-sector configuration can be implemented using existing antenna hardware and controlled at slow time scales, making it compatible with current NR scheduling and emerging radio access network (RAN) control architecture. To further protect UAV links under strong inter-cell interference, we incorporate an NR-compatible time-domain interference coordination (TDIC) mechanism. 
Within this framework, we formulate a max–min SIR optimization problem over a spatial grid of UAV locations. 

The optimization problem is highly non-convex and NP-hard due to the strong coupling between antenna tilt configurations and multi-cell interference. To solve this, we propose a hybrid genetic algorithm (GA) that combines global evolutionary search with deterministic local refinement and a particle swarm optimization (PSO) to reduce interference under both uncoordinated (US) and coordinated (CS) slots.

The key contributions of this study are as follows: 
\begin{itemize}
\item Given the growing interest in 3D UAV users, we propose a fundamentally different approach where each BS is equipped with a complementary uptilted antenna sector to serve aerial UAV users, while the preexisting downtilted antenna is designed to serve the ground users. Such an approach improves worst-case UAV downlink SIR in multi-cell 5G NR macro networks without modifying the NR air interface or UAV hardware.
\item For considered cellular networks, addressing inter-cell interference at UAVs in an NR-compliant manner requires explicit coordination across BSs. We cast a max-min SIR optimization problem for the downlink of the involved UAV over the coordinated uptilt angles of the BSs, focusing on mitigating interference-limited links within the area of interest.
\item  We design and analyze a hybrid GA approach that performs better compared to a baseline GA technique and offers reliable worst-case SIR gains under high inter-cell interference scenarios.
\item We present a comprehensive evaluation of the benefits of coordinated uptilt optimization and NR-compatible TDIC through different approaches via multi-cell simulations.
\end{itemize}

The remainder of this paper is structured as follows. We describe the system model and problem formulation, including multi-cell network topology, UAV deployment scenario, intercell interference coordination approach, antenna configuration, and the max-min SIR optimization objective in Section \ref{sec:system_model}. Section~\ref{sec:uptilt_angle_opt_approach} discusses the uptilt angle optimization methods. In Section~\ref{sec:results_discussion}, we describe the simulation setup and the results. Finally, Section~\ref{sec:conclusion} summarizes the conclusion and future directions.

\section{System Model and Problem Formulation}
\label{sec:system_model}
\subsection{Network Architecture and Assumption}
\begin{figure}[t]
    \centering        \includegraphics[width=3.5in,trim=.2in 1.2in 0in 1.7in,clip]
    {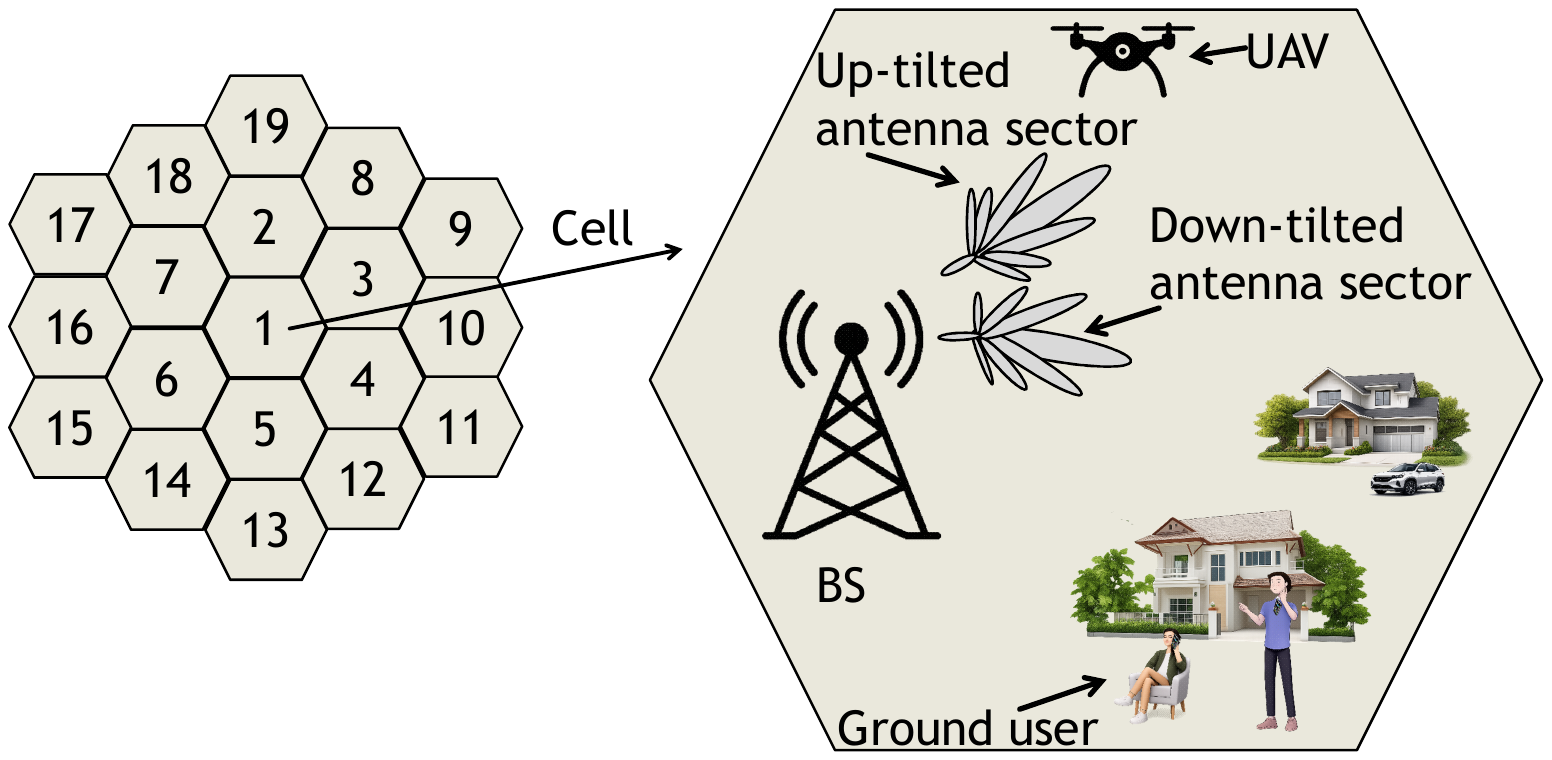}    
    \caption{Multi-cell macro network with co-located downtilted and uptilted antenna sectors serving ground and aerial users.}
    \label{fig:sys_model}    
    \vspace{-.2in}
\end{figure}

We consider the downlink of a multi-cell 5G NR macro network serving both terrestrial and aerial users \cite{TS22125}, or UAVs, as illustrated in Fig.~\ref{fig:sys_model}. A standard 19-cell hexagonal layout with a wrap-around model is assumed, where each cell is served by a single BS. Each BS is equipped with two co-located vertical antenna sectors, which we denote as the booster cell, where a downtilted sector serves ground users for good coverage and an additional uptilted sector provides aerial coverage by steering energy toward UAVs operating at altitude. The downtilt angle is fixed, whereas the uptilt angle of each BS is configurable and can be optimized. As shown in Fig.~\ref{fig:sys_model}, as an example, the UAV is associated with the center cell, and the surrounding cells act as a dominant source of inter-cell interference.

Here, the UAVs are considered to be quasi-static within the tilt optimization interval, and their horizontal positions are modeled by a coarse spatial grid with a constant height. Second, the network has complete knowledge of the UAV channel state information at the network to facilitate coverage-aware interference coordination. Third, all BSs transmit with identical power and operate on the same carrier frequency. The UAV association is based on the highest average received power, with each UAV served by a single BS at any given time. Fourth, uplink transmission, power control, and beamforming are not considered and are left for future studies. 

\subsection{UAV Deployment and Association}
We focus on UAVs located in the sky above the center cell. To capture spatial variability, we discretize the horizontal region inside the center hexagon into a grid with spacing $\Delta_g$ = $10$ m. Let $u = \{1,\ldots,U\}$ denote the set of grid points that lie inside the center hexagon with horizontal coordinates $\mathbf{r}_u = (x_u,y_u)$. All UAVs are placed at the same altitude $h$ above the ground. Hence, the three-dimensional position of the UAV $u$ is: $\tilde{\mathbf{u}}_u = (x_u,y_u,h)$. 

Let $P^\mathrm{ut}_{b,u}$ and $P^\mathrm{dt}_{b,u}$ denote the average received power at the UAV $u$ from the uptilted and downtilted arrays of the ground BS $b$, respectively. The composite received power is:
\begin{equation}
    P^{\text{tot}}_{b,u} = \max\left(P^\mathrm{ut}_{b,u}, P^\mathrm{dt}_{b,u}\right)~,
    \label{eq:received_power}
\end{equation}
and the serving BS index for UAV $u$ is:
\begin{equation}
    b_u^\star = \arg \max_{b \in \mathcal{B}} P^{\text{tot}}_{b,u}
    \label{eq:assoc}
\end{equation}

\subsection{3D Antenna Model}
We adopt a standard 3GPP vertical element pattern for both downtilted and uptilted arrays \cite{3GPP38901}. Let $\theta$ denote the elevation angle measured from the horizontal plane (positive when pointing upward), and let $\theta^\circ$ be the angle in degrees. The vertical element gain in dB at $\theta^\circ$ is:
\begin{equation}
    G_e(\theta^\circ) = G_{e,\max} - \min\left( 12 \left(\frac{\theta^\circ}{\theta_{3\text{dB}}}\right)^2,~\mathrm{SLA}_v \right)~,
\end{equation}
where $G_{e,\max}$ is the maximum element gain, $\theta_{3\text{dB}}$ is the 3 dB beamwidth, and $\mathrm{SLA}_v$ is the side-lobe attenuation in the vertical direction.

Each vertical array at BS $b$ is modeled as a uniform linear array (ULA) with $N_t$ elements and half-wavelength spacing. The array is steered toward an electrical tilt angle $\phi$, where $\phi = \phi_\mathrm{dt}$ for the downtilted array and $\phi = \phi_\mathrm{ut}$ for the uptilted array. For a given $\theta$, the normalized array factor is:
\begin{equation}
A(\theta,\phi) = \frac{1}{\sqrt{N_t}} \frac{\sin\left(\tfrac{N_t\pi}{2}(\sin\theta - \sin\phi)\right)}{\sin\left(\tfrac{\pi}{2}(\sin\theta - \sin\phi)\right)}~,
\end{equation}
and the corresponding array gain in dB is:
\begin{equation}
G_a(\theta,\phi) = 10 \log_{10} \left|A(\theta,\phi)\right|^2
\end{equation}
The composite vertical gain is:
\begin{equation}
G(\theta,\phi) = G_e(\theta^\circ) + G_a(\theta,\phi)~, \quad \theta^\circ = \theta \cdot \frac{180}{\pi}
\end{equation}
For both uptilted and downtilted antenna arrays, the elevation angle, $\theta_{b,u}$, is computed from the same BS–UAV geometry, while the corresponding antenna gains differ only in the applied electrical tilt angle.
For the uptilted and downtilted antenna array, the composite antenna gain between BS and UAV is given by:
\begin{equation}
G^{\mathrm{ut}}_{b,u}
= G\!\left(\theta_{b,u}, \phi_{\mathrm{ut},b}\right)
= G_e\!\left(\theta_{b,u}^{\circ}\right)
+ G_a\!\left(\theta_{b,u}, \phi_{\mathrm{ut}}\right)
\label{eq:gain_ut}
\end{equation}
\begin{equation}
G^{\mathrm{dt}}_{b,u}
= G\!\left(\theta_{b,u}, \phi_\mathrm{dt}\right)
= G_e\!\left(\theta_{b,u}^{\circ}\right)
+ G_a\!\left(\theta_{b,u}, \phi_\mathrm{dt}\right)
\label{eq:gain_dt}
\end{equation}

\subsection{3D Propagation Model}
The horizontal 2D distance between the BS site and the UAV projection is given by:
\begin{equation}
d_{b,u}^\mathrm{2D} = \sqrt{(x_u - x_b)^2 + (y_u - y_b)^2}~,
\end{equation}

Let $h_b$ denote the BS height. The vertical height differences for uptilt and downtilt sectors, where $h_b$ = $h_{b,\mathrm{dt}}$, $h_{b,\mathrm{ut}}$ = $h_{b,\mathrm{dt}}+1$, the spacing between $h_{b,\mathrm{ut}}$ and $h_{b,\mathrm{dt}}$ is 1 m,  are:
\begin{equation}
\Delta h_\mathrm{ut} = h - h_{b,\mathrm{ut}}, \quad \Delta h_\mathrm{dt} = h - h_{b,\mathrm{dt}}~,
\end{equation}
The three-dimensional distances ($\ell$) and elevation angles are:
\begin{align}
\ell^\mathrm{ut}_{b,u} = \sqrt{(d_{b,u}^{\mathrm{2D}})^2 + (\Delta h_\mathrm{ut})^2},
\theta^\mathrm{ut}_{b,u}=\tan^{-1}\left(\frac{\Delta h_\mathrm{ut}}{d_{b,u}^{\mathrm{2D}}}\right),
\end{align}
\begin{align}
\ell^\mathrm{dt}_{b,u} = \sqrt{(d_{b,u}^{\mathrm{2D}})^2 + (\Delta h_\mathrm{dt})^2},
\theta^\mathrm{dt}_{b,u} = \tan^{-1}\left(\frac{\Delta h_\mathrm{dt}}{d_{b,u}^{\mathrm{2D}}}\right),
\end{align}

\begin{figure}[t]
    \centering
    \includegraphics[width=\columnwidth,trim=0in 1.3in 0in 1.2in, clip]{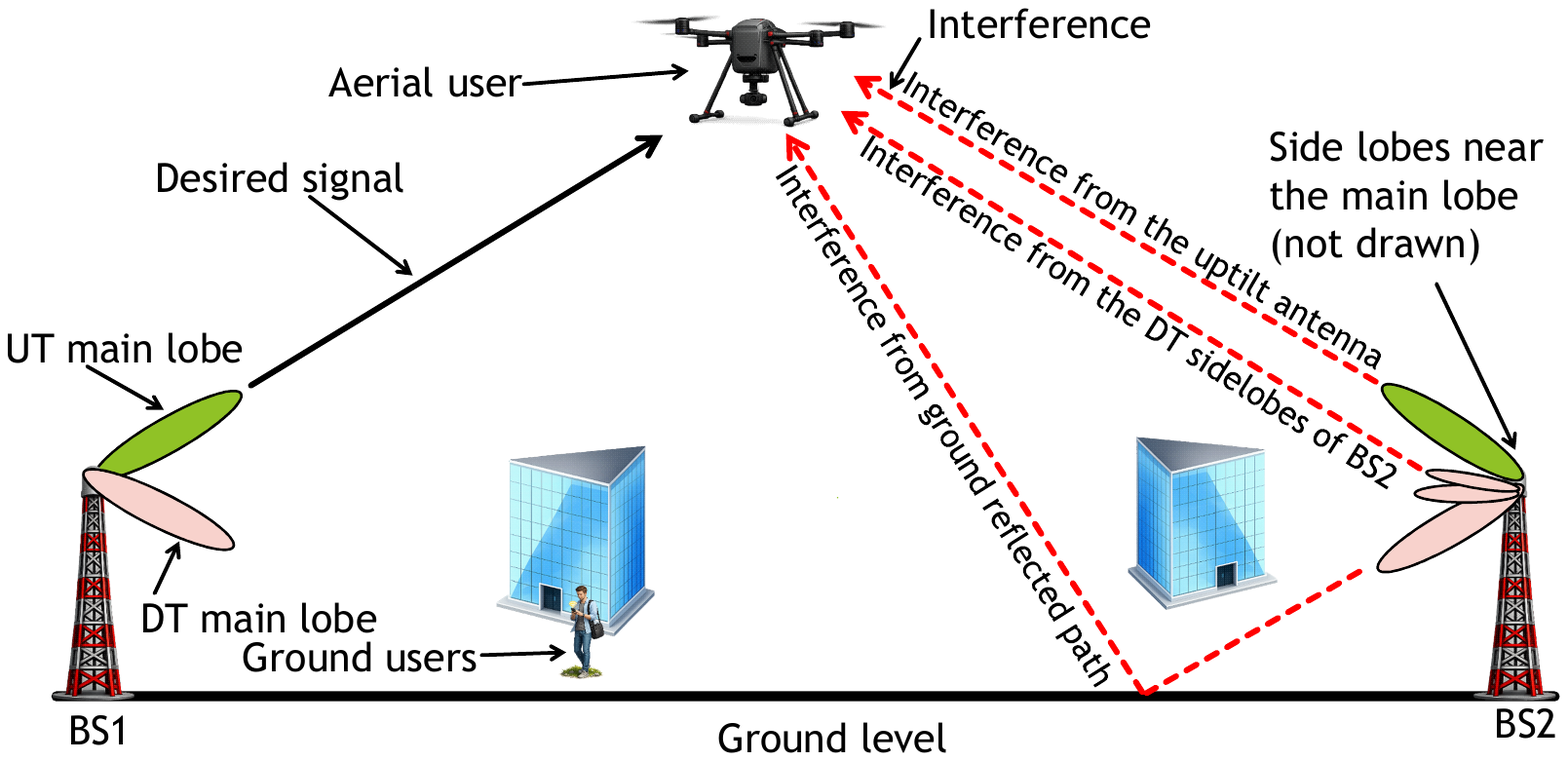}
    \caption{Illustration of the dominant downlink signal and interference components for an aerial user. The desired signal is received from the serving BS through the main lobe of the uptilted (UT) antenna, while strong interference arises from neighboring BSs via DT sidelobes and ground-reflected paths associated with DT.}
    \label{fig:ground_reflection}
    \vspace{-.2in}
\end{figure}

We consider a height-dependent propagation model that captures the effect of ground reflection as shown in Fig.~\ref{fig:ground_reflection}. 
The effective path loss exponent for a UAV at $h$ is:
\begin{equation}
\alpha(h) =
\begin{cases}
\alpha_0 - \dfrac{h}{h_\mathrm{b}}(\alpha_0 - 2), & h < 2h_\mathrm{b} \\ 
2, & h \ge 2h_\mathrm{b}~,
\end{cases}
\end{equation}

The downtilted array experiences a strong ground reflection (GR) that contributes an additional path between the BS and the UAV. Following a two-ray style model illustrated in Fig.~\ref{fig:ground_reflection}, the received power is divided into direct and reflected components:
\begin{align}
P^{\text{dir}}_{b,u} &= P_\mathrm{b} \left(\frac{\lambda}{4\pi}\right)^2 \frac{G(\theta^{\mathrm{dt}}_{b,u},\phi_{\mathrm{dt}})}{\left(\ell^{\mathrm{dt}}_{b,u}\right)^{\alpha(h)}},\\
P^{\text{ref}}_{b,u} &= P_b \left(\frac{\lambda}{4\pi}\right)^2
\frac{|R_g|^2 G_e^{r}(h)}{(r_{1,b,u} + r_{2,b,u})^{\alpha(h)}}~,
\end{align}

where $R_g$ is an effective Fresnel reflection coefficient depending on the incidence angle and ground relative permittivity, $G_e^{r}(h)$ is a height-dependent effective element gain for the reflected path, and $r_{1,b,u}$, $r_{2,b,u}$ are the distances from BS to ground intercept and from ground to UAV, respectively. Then, the total received power from the downtilted array is:
\begin{equation}
P^{\mathrm{dt}}_{b,u} = P^{\text{dir}}_{b,u} + P^{\text{ref}}_{b,u}~,
\label{eq:receied_power_from_dt_sector}
\end{equation}
The uptilted array does not have a significant GR toward UAVs since the main beam points upward. We therefore model the received power from the uptilted array as:
\begin{equation}
P^{\mathrm{ut}}_{b,u} = P_{\text{b}} \left(\frac{\lambda}{4\pi}\right)^2
\frac{G(\theta^{\mathrm{ut}}_{b,u},\phi_{\mathrm{ut},b})}{\left(\ell^{\mathrm{ut}}_{b,u}\right)^{\alpha(h)}}~,
\label{eq:receied_power_from_ut_sector}
\end{equation}

Substituting \eqref{eq:receied_power_from_dt_sector} and \eqref{eq:receied_power_from_ut_sector} into \eqref{eq:received_power}, serving BS $b_u^\star$ is determined from \eqref{eq:assoc}, which provides the maximum average received power to the UAV.
\subsection{SIR and TDIC-Based Interference Coordination}
Due to the dominant LoS exposure between UAVs and multiple BSs, aerial downlink links in macrocellular networks are primarily interference-limited \cite{donggu_lee}. We verified that including thermal noise does not alter the relative performance trends, since interference remains dominant across the considered scenarios. Hence, thermal noise is neglected in the following analysis.

Fig.~\ref{fig:ground_reflection} illustrates the dominant desired signal and interference components experienced by a UAV in a multi-cell deployment. While the desired signal is received through the main lobe of the UT antenna of the serving BS, strong interference arises from DT sidelobes of neighboring BSs as well as from ground-reflected paths associated with DT transmissions. 

\begin{figure}[t]
    \centering        \includegraphics[width=3.5in,trim=.65in 3.35in 0.3in 3in,clip]
    {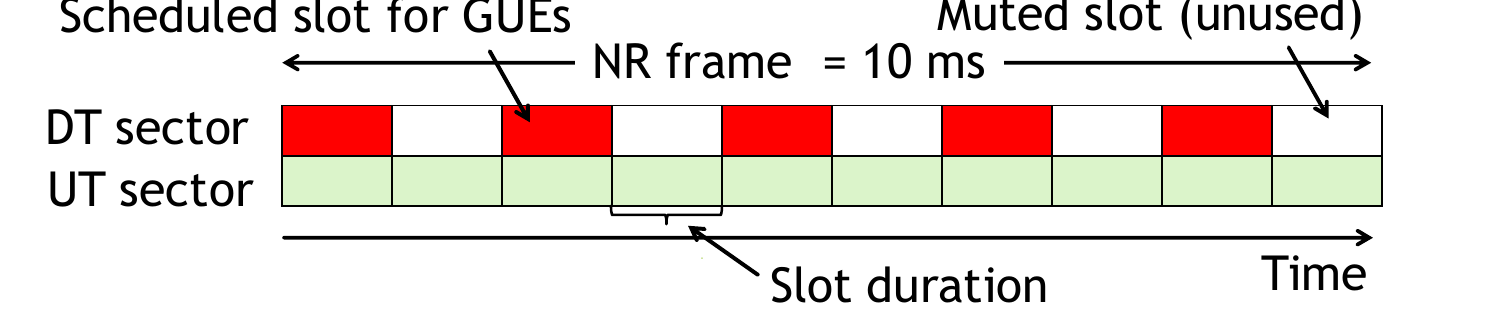}    
    \caption{NR-compatible slot-level muting between DT and UT antenna sectors.}
    \label{fig:dt_ut_slot}
    \vspace{-.2in}
\end{figure}

To mitigate the severe inter-cell interference experienced by UAVs, we consider the TDIC mechanism shown in Fig.~\ref{fig:dt_ut_slot} that selectively mutes DT data transmissions during coordinated slots (CS), while preserving essential control signaling for network operation. By muting DT data transmissions during CS, the proposed TDIC mechanism significantly reduces the aggregate interference level at the UAV receiver.

In uncoordinated slots (US), both UT and DT antenna sectors transmit simultaneously. Hence, in the US slot, UAV experiences aggregate interference from the UT and DT sectors of all non-serving BSs, as well as from the sidelobes of the DT sector of its serving BS, as illustrated in Fig.~\ref{fig:ground_reflection}. The downlink SIR at the UAV $u$ during US is given by:
\begin{equation}
\gamma^{\mathrm{US}}_{u} =
\frac{P^{\mathrm{ut}}_{b_u^\star,u}}
{\sum\limits_{b \in \mathcal{B},\, b \neq b_u^\star}
\left(P^{\mathrm{ut}}_{b,u} + P^{\mathrm{dt}}_{b,u}\right)
+ P^{\mathrm{dt}}_{b_u^\star,u}}
\label{eq:sir_us}
\end{equation}

 During CS, UAVs are served primarily through UT antenna sectors, and DT-induced interference is effectively suppressed. Accordingly, the downlink SIR at UAV $u$ during CS is:
\begin{equation}
\gamma^{\mathrm{CS}}_{u} =
\frac{P^{\mathrm{ut}}_{b_u^\star,u}}
{\sum\limits_{b \in \mathcal{B},\, b \neq b_u^\star}
P^{\mathrm{ut}}_{b,u}}
\label{eq:sir_cs}
\end{equation}

We consider a duty-cycle parameter, $\beta$, to maintain the proportion of coordinated slots within an NR frame. It controls the trade-off between UAV interference mitigation and ground user equipment (GUE) quality of service (QoS) continuity. 

\subsection{Problem Formulation}
Our objective is to improve the worst-case downlink reliability of UAVs in a multi-cell 5G NR macro network. Since US slots represent the most interference-limited operating condition, we focus on maximizing the minimum UAV downlink SIR in the US across the spatial grid of UAV locations. Let $\boldsymbol{\phi}_\mathrm{ut} = [\phi_{\mathrm {ut},1}, \ldots, \phi_{\mathrm{ut},|\mathcal{B}|}]^T$ denote the vector of UT antenna tilt angles, where each element corresponds to the UT tilt angle of a BS. The optimization problem is formulated as:

\begin{equation}
\begin{aligned}
\max_{\boldsymbol{\phi}_{\mathrm{ut}}} \quad 
& \min_{u \in \mathcal{U}} \; 10\log_{10}\!\left(\gamma_u^{\mathrm{US}}(\boldsymbol{\phi}_{\mathrm{ut}})\right) \\
\text{s.t.} \quad 
& \phi_{\min} \le \phi_{\mathrm{ut},b} \le \phi_{\max}, \quad \forall b \in \mathcal{B}
\end{aligned}
\label{eq:opt}
\end{equation}
In \eqref{eq:opt}, the UT tilt angles are jointly optimized across all BSs under US, which represents the worst-case interference condition for UAV links. Using the optimized UT tilts, we subsequently evaluate UAV performance in CS slots, where DT data transmissions are muted, to quantify the additional interference mitigation benefits provided by TDIC. 
Due to the strong coupling between UT antenna tilt configurations and multi-cell interference, the optimization problem in~\eqref{eq:opt} is highly non-convex and does not admit a closed-form solution. Hence, we consider evolutionary algorithms in Section~\ref{sec:uptilt_angle_opt_approach}.

\section{Uptilt Angle Optimization}
\label{sec:uptilt_angle_opt_approach}
\subsection{Hybrid Genetic Algorithm}
We propose a hybrid genetic algorithm (GA) to solve the max–min SIR optimization problem by integrating global evolutionary search techniques with deterministic local refinement. In this approach, each candidate solution represents a complete set of BS uptilt angles. The GA explores the search space using roulette-wheel selection, single-point crossover, mutation, and elitist replacement \cite{GA9002372}, where fitness is defined as the negative of the minimum SIR of the UAVs over the spatial grid. The GA selects parent solutions using roulette-wheel selection, where better candidates are chosen with higher probability. In crossover, elements of two candidate solutions combine to create a new one, whereas mutation is a small change to avoid getting stuck with a bad solution and to increase the possibilities. In elitist selection, the best solutions obtained from one step of the algorithm are carried over to the next step so that they will not be lost. 

To improve convergence and fine-tune the solution, the best GA result is further refined using a coordinate-wise local search (LS). Each uptilt angle is adjusted by $\pm \Delta$ degrees and accepted if the fitness value increases and the step size decreases until convergence, yielding small but consistent gains in worst-case SIR.

\subsection{Particle Swarm Optimization}
In PSO, each particle corresponds to a solution vector for the BS uptilt angles. Particles refine their solutions following velocity dynamics with a combination of inertia, personal experience, and global knowledge. In promoting convergence and physically valid solutions, particle velocity and uptilt angle values are limited. The particle swarm starts with a nominal tilt solution with random variations for efficient search space investigation and adequate solution generation. A localized search is also employed with a global optimization solution via PSO for a final improved solution with a supporting SIR.

\section{Numerical Results}
\label{sec:results_discussion}
In this section, we present the simulation results based on the parameters shown in Table~\ref{tab:sim_params}, and evaluate the performance of coordinated uptilt optimization under varying network densities and antenna configurations.

\begin{table}[t]
\centering
\caption{Simulation parameters.}
\label{tab:sim_params}
\begin{tabular}{ll}
\toprule
Parameter & Value \\ \midrule
Carrier frequency $f_c$ & \SI{3.5}{GHz} \\
BS transmit power $P_{\text{b}}$ & \SI{46}{dBm} \\
BS antenna heights (DT / UT) $(h_\mathrm{dt}, h_\mathrm{ut})$ & \SI{30}{m}, \SI{31}{m} \\
Vertical separation between\\ DT and UT sectors $h_{\text{sep}}$ & \SI{1}{m} \\
Inter-site distance $\mathrm{ISD}$ & \SI{500}{m}, \SI{1000}{m} \\
UAV heights $h$ & $\{100, 200\}$ m \\
Element max gain $G_{e,\max}$ & \SI{8}{dBi} \\
Vertical 3 dB beamwidth $\theta_{3\text{dB}}$ & $65^\circ$ \\
Side-lobe attenuation (vertical) $SLA_v$ & \SI{30}{dB} \\
Number of vertical elements $N_t$ & $8$ ($4$ and $16$) \\
Down-tilt angle $\phi_{dt}$ & $-6^\circ$ \\
UT tilt bounds $[\phi_{\min},\phi_{\max}]$ & $[5^\circ,89^\circ]$ \\
Medium dry ground permittivity $\varepsilon_r$~\cite{3GPP38901}\cite{permititvity6214705}& $15$\\
Path-loss exponent $\alpha_0$ & $3.5$ \\
UAV grid spacing $\Delta_g$ & \SI{10}{m} \\
GA population size $M$ & $200$ \\
Number of GA generations $I_{\max}$ & $100$ \\
Mutation probability $p_{\text{mut}}$ & $0.1$ \\
Initial/minimum step size of Hybrid GA & $2^\circ \rightarrow 0.1^\circ$ \\
Hybrid local iterations & up to $50$ \\
PSO swarm size, iterations & $200$ particles, $100$ \\
PSO inertia weight $w$ & $0.72$ \\
PSO cognitive/social coefficients $(c_1,c_2)$ & $1.45, 1.45$ \\
PSO velocity limit & $8^\circ$/iteration \\
\bottomrule
\end{tabular}
\end{table}

\subsection{Spatial Coordination and Network Density Effects}
\label{sec:isd_alt}
\begin{figure}[t]
\centering
\begin{subfigure}[t]{0.24\textwidth}
  \centering
  \includegraphics[width=\linewidth,trim=0in 1in 0in 1.8in,clip]{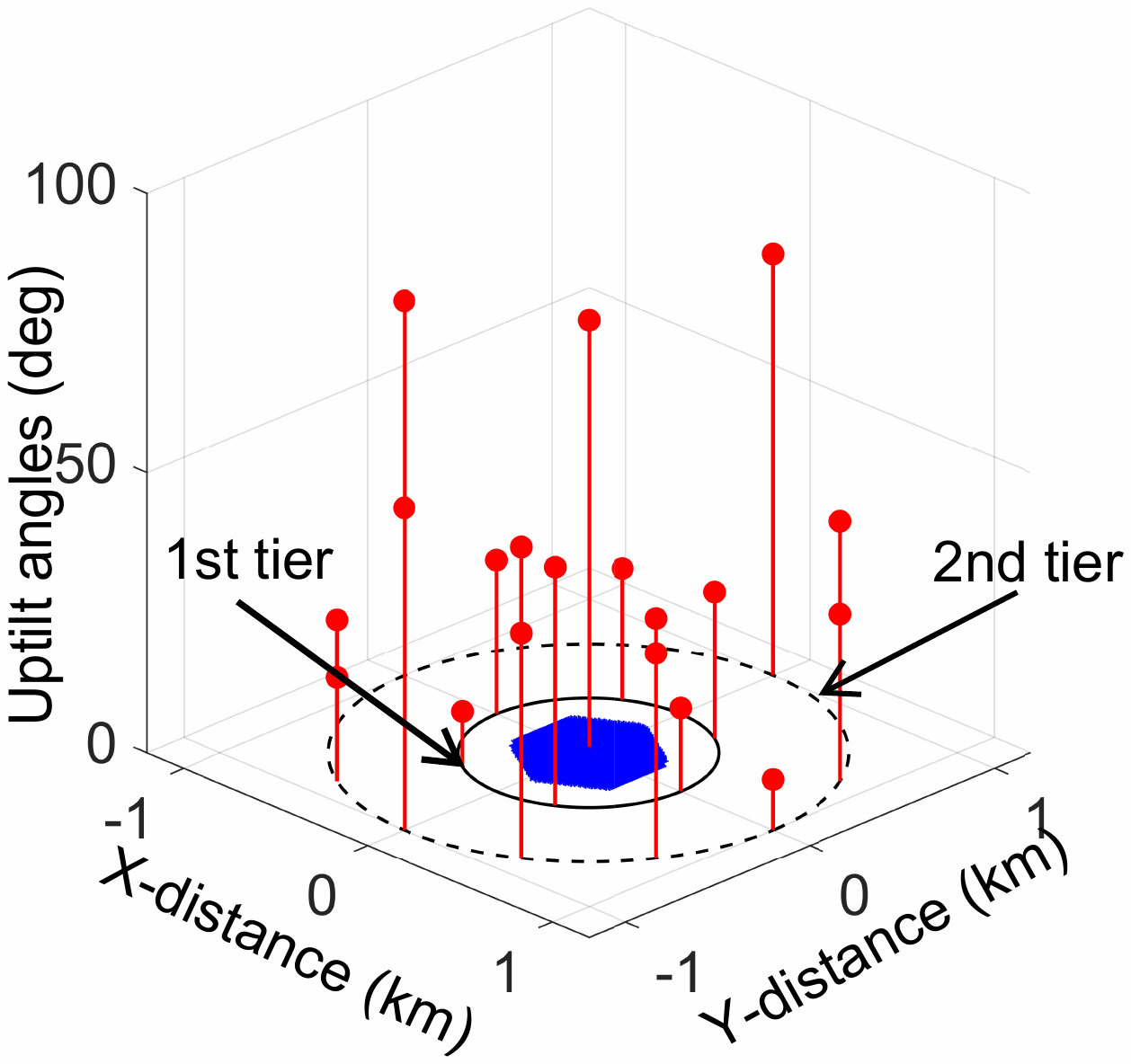}
  \caption{Random}
\end{subfigure}
\begin{subfigure}[t]{0.24\textwidth}
  \centering
  \includegraphics[width=\linewidth,trim=0in 1in 0in 1.8in,clip]{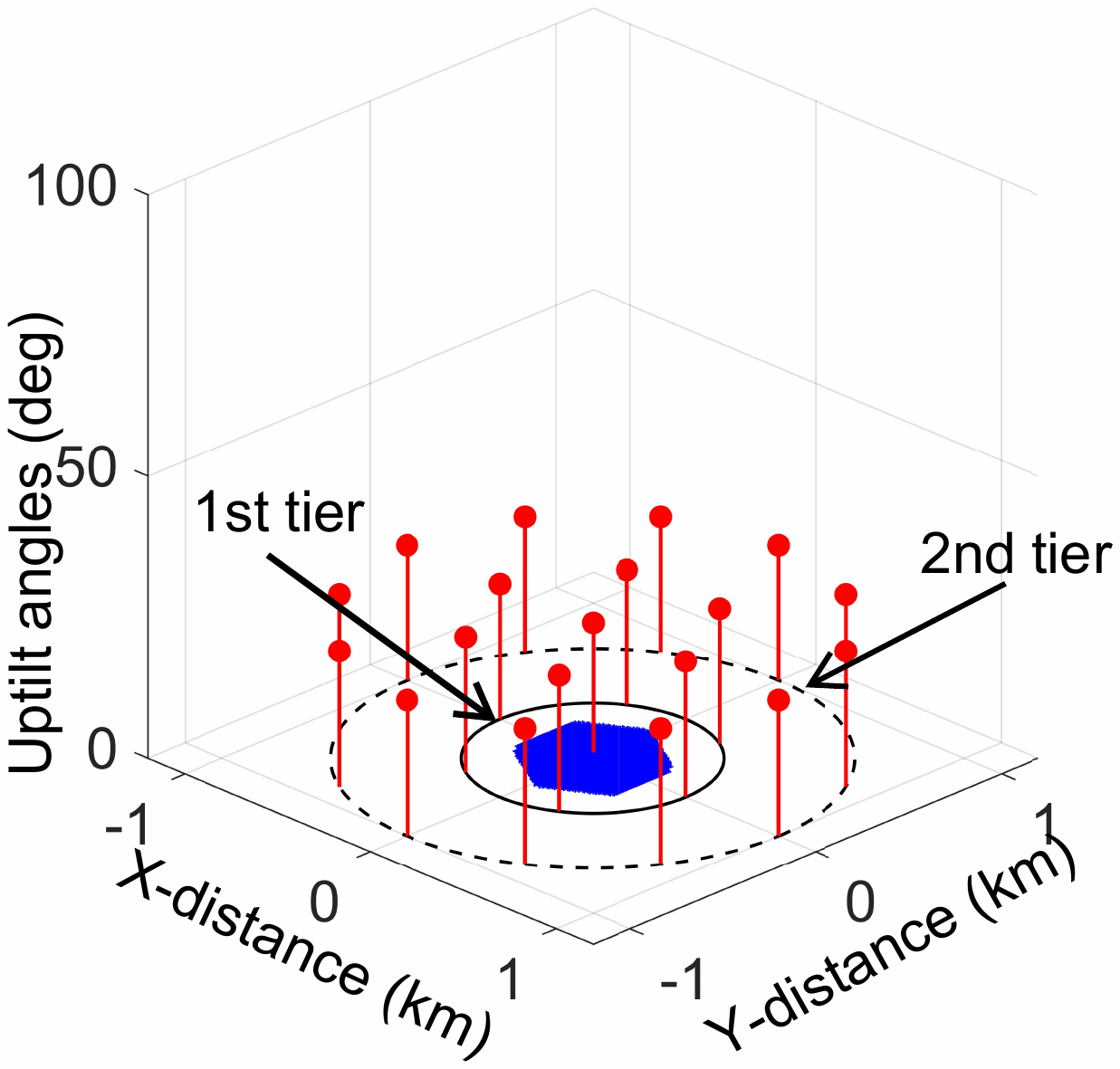}
  \caption{Single}
\end{subfigure}
\begin{subfigure}[t]{0.24\textwidth}
  \centering
  \includegraphics[width=\linewidth,trim=0in 1in 0in 1.8in,clip]{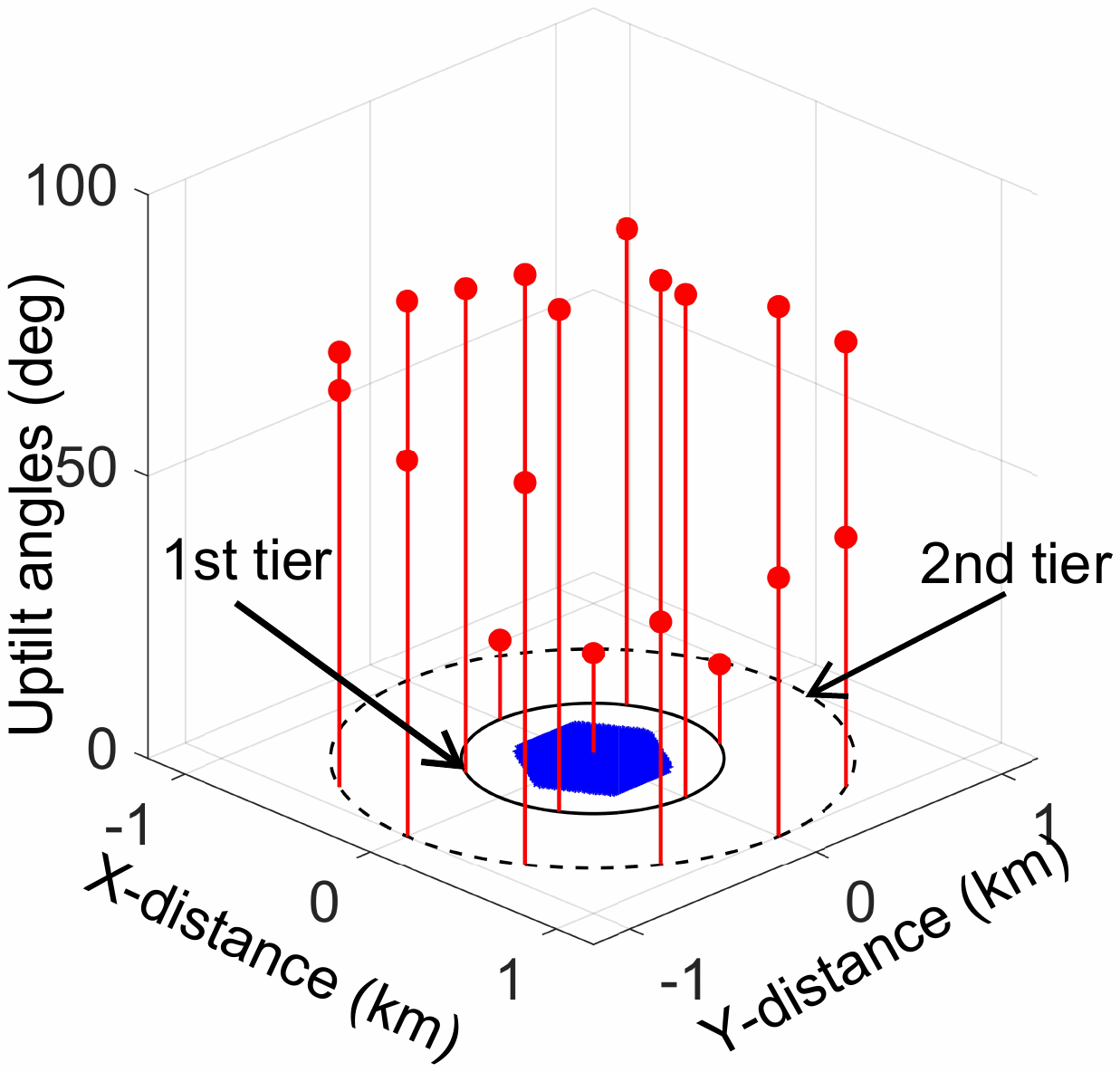}
  \caption{HGA}
\end{subfigure}
\begin{subfigure}[t]{0.24\textwidth}
  \centering
  \includegraphics[width=\linewidth,trim=0in 1in 0in 1.8in,clip]{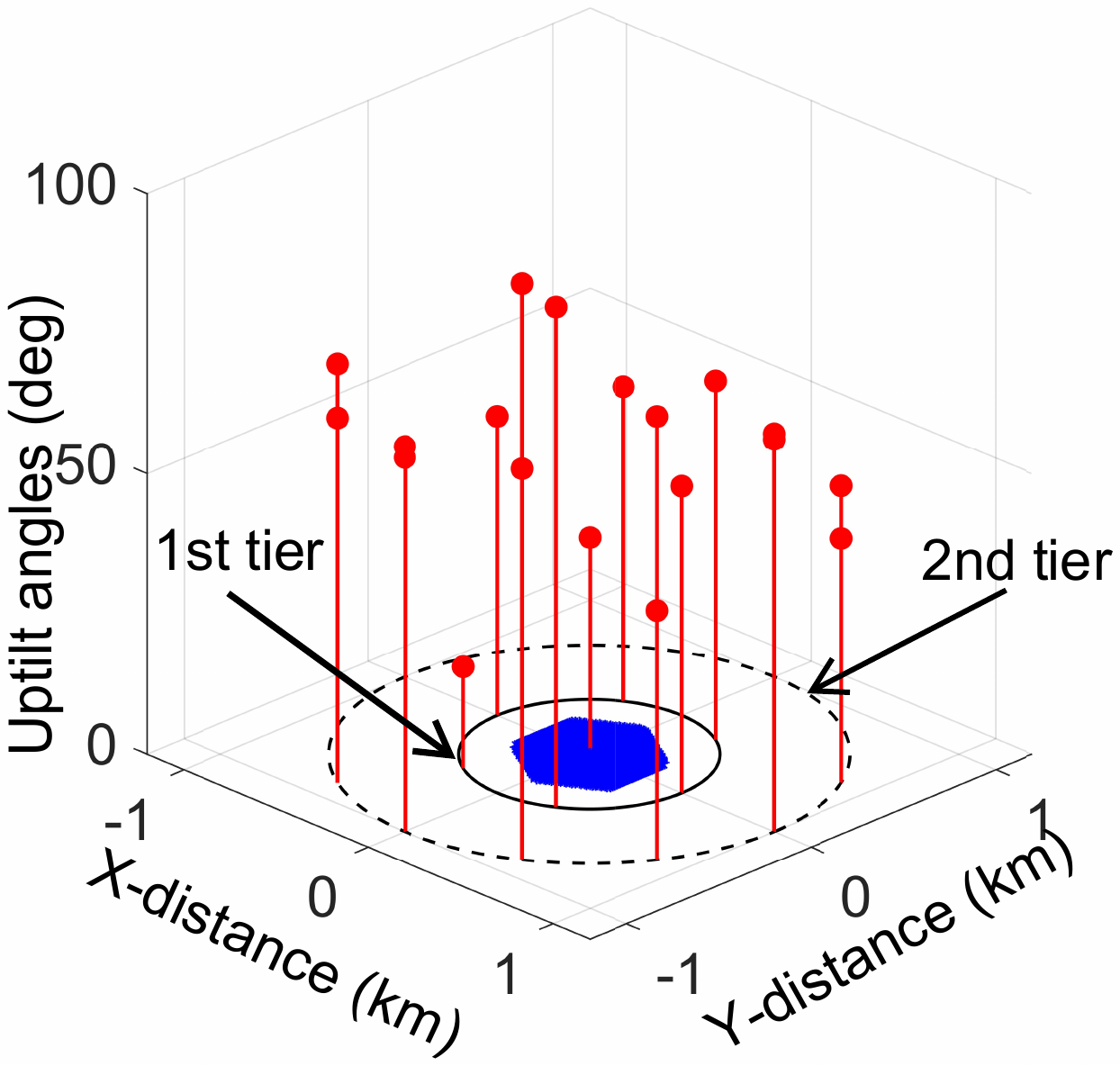}
  \caption{PSO}
\end{subfigure}
\caption{Uptilt angle selection for ISD = 500 m, $h$ = 200 m for different approaches.}
\vspace{-.1in}
\label{fig:optimal_uptilt}
\end{figure}

Fig.~\ref{fig:optimal_uptilt} shows results for uptilt angle selection. Fig.~\ref{fig:optimal_uptilt}(a) shows the random approach, where we see that the UT angles do not show any significant pattern over the network. In this method, the serving and interfering cells use arbitrary UT angles. As a result, the main antenna beam is not aligned with the UAV, and unnecessary radiation is sent towards the sky, which causes interference. In contrast, in the single-angle approach, all BSs use the same uniform value for the UT. The resulting tilt pattern is flat and independent of the geometry shown in Fig.~\ref{fig:optimal_uptilt}(b). Although this approach provides some gain in the upward direction for the serving link, the elevation change with distance between the BSs and the UAV is not utilized effectively. Hence, the interference from the outer tiers has not been mitigated effectively. In the case of the HGA-based solution, we observe a well-structured spatial distribution in Fig.~\ref{fig:optimal_uptilt}(c), where uptilts with the highest values are considered near the serving cell, and they decrease gradually with the increase of the distance. 
Finally, the PSO solution results in a spatial distribution of optimized uptilt angles across cells, as shown in Fig.~\ref{fig:optimal_uptilt}(d).

\begin{figure*}[t]
\centering
\begin{subfigure}[t]{0.34\textwidth}
  \centering
  \includegraphics[width=\linewidth]{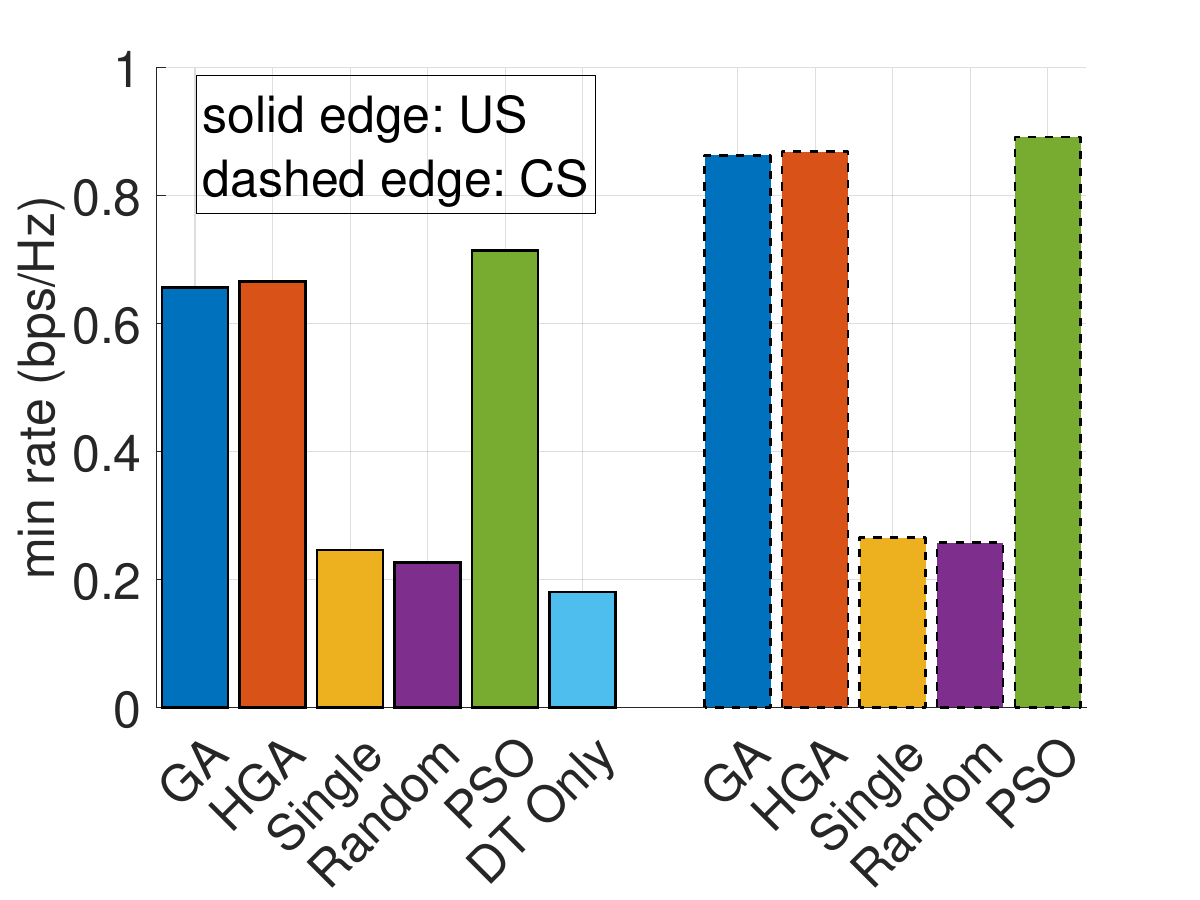}
  \caption{min rate}
\end{subfigure}
\begin{subfigure}[t]{0.34\textwidth}
  \centering
  \includegraphics[width=\linewidth]{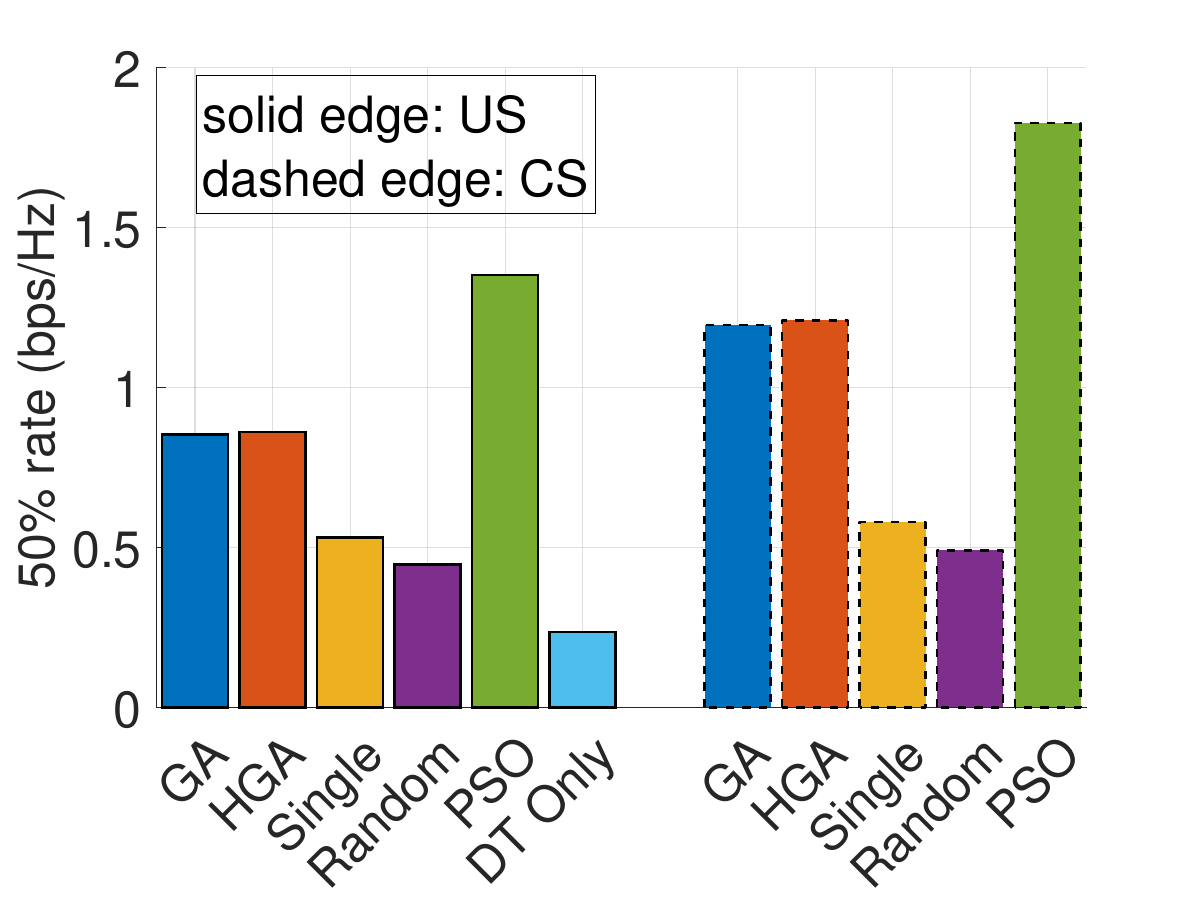}
  \caption{50\% rate}
\end{subfigure}
\begin{subfigure}[t]{0.34\textwidth}
  \centering
  \includegraphics[width=\linewidth]{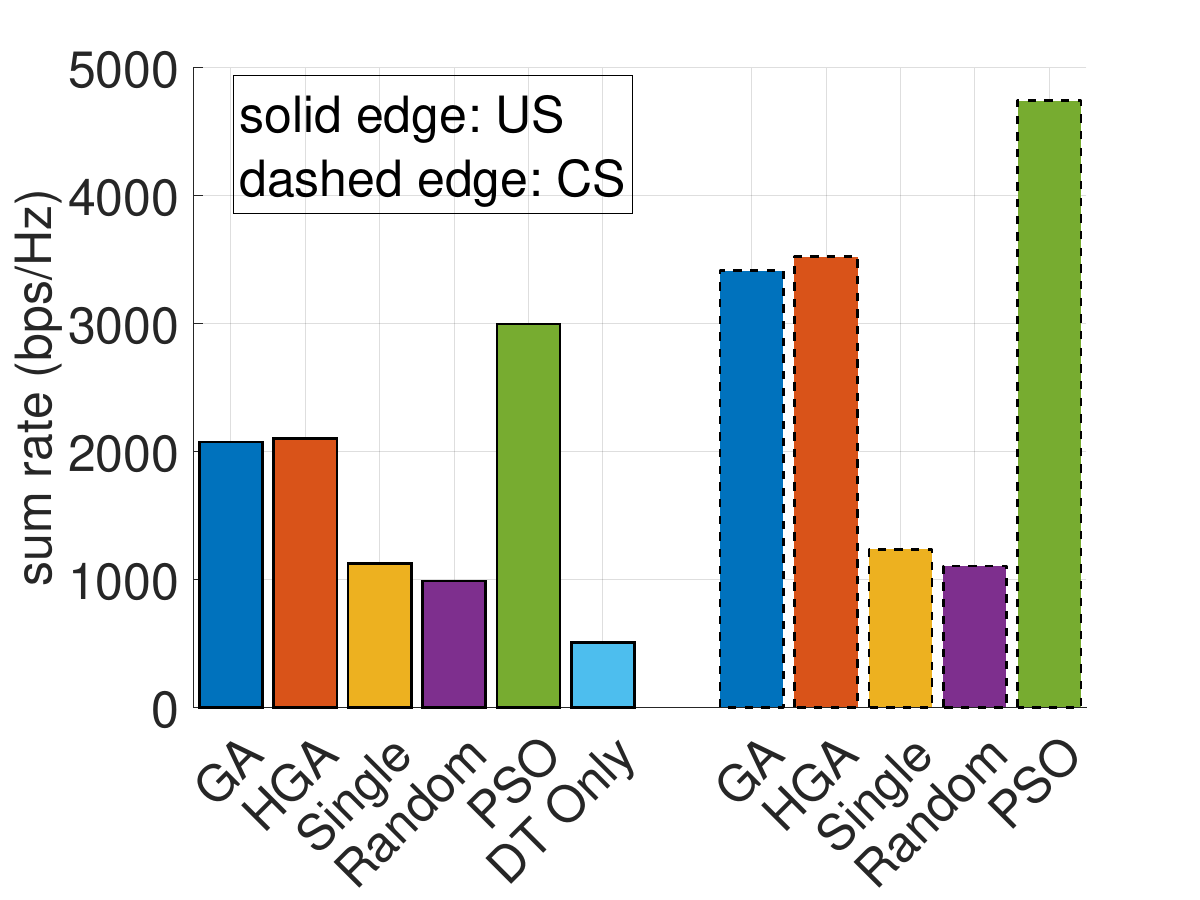}
  \caption{sum rate}
\end{subfigure}
\caption{Rate with ISD = 500 m and $h$ = 200 m for US and CS subframes considering different algorithms.}
\label{fig:rate_500m_200m}
\vspace{-.2in}
\end{figure*}

\begin{figure*}[t]
\centering
\begin{subfigure}[t]{0.34\textwidth}
  \centering
  \includegraphics[width=\linewidth]{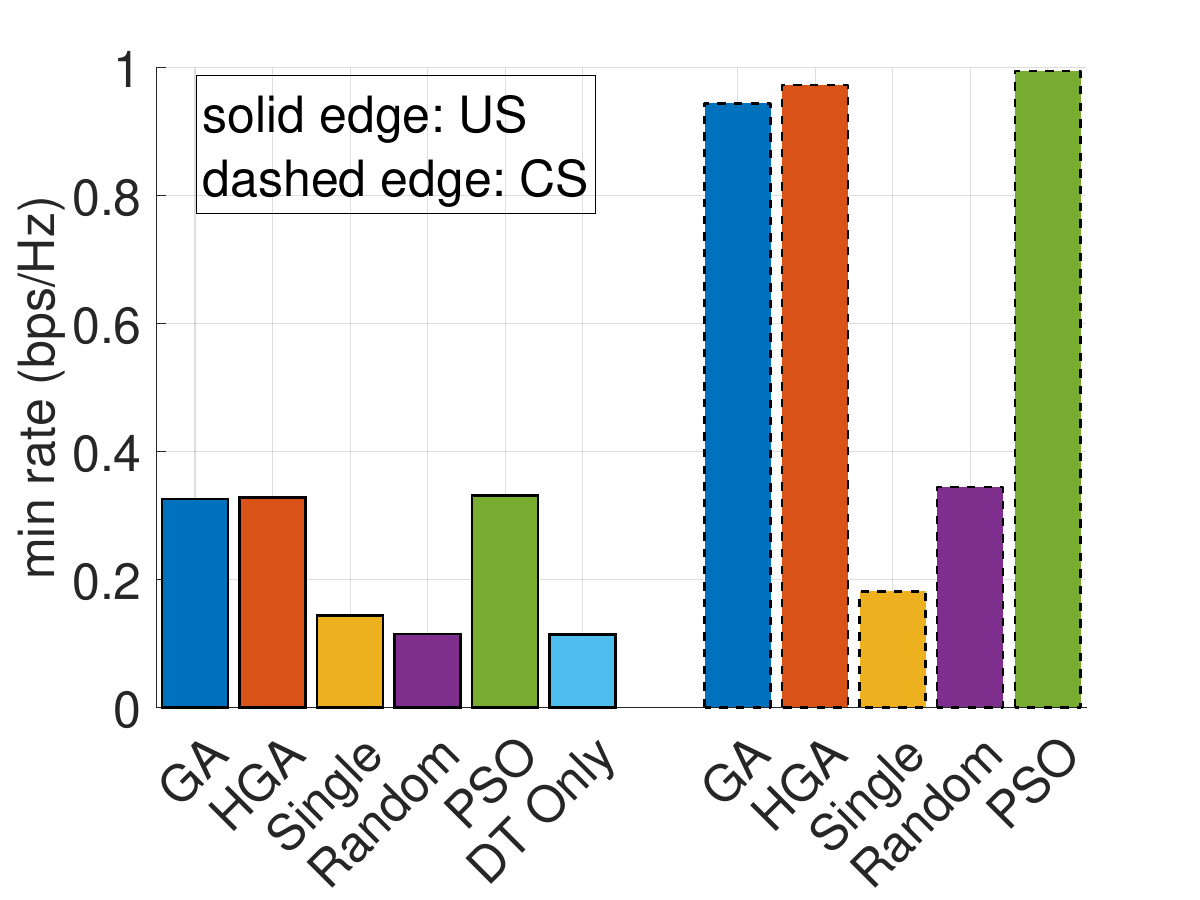}
  \caption{min rate}
\end{subfigure}
\begin{subfigure}[t]{0.34\textwidth}
  \centering
  \includegraphics[width=\linewidth]{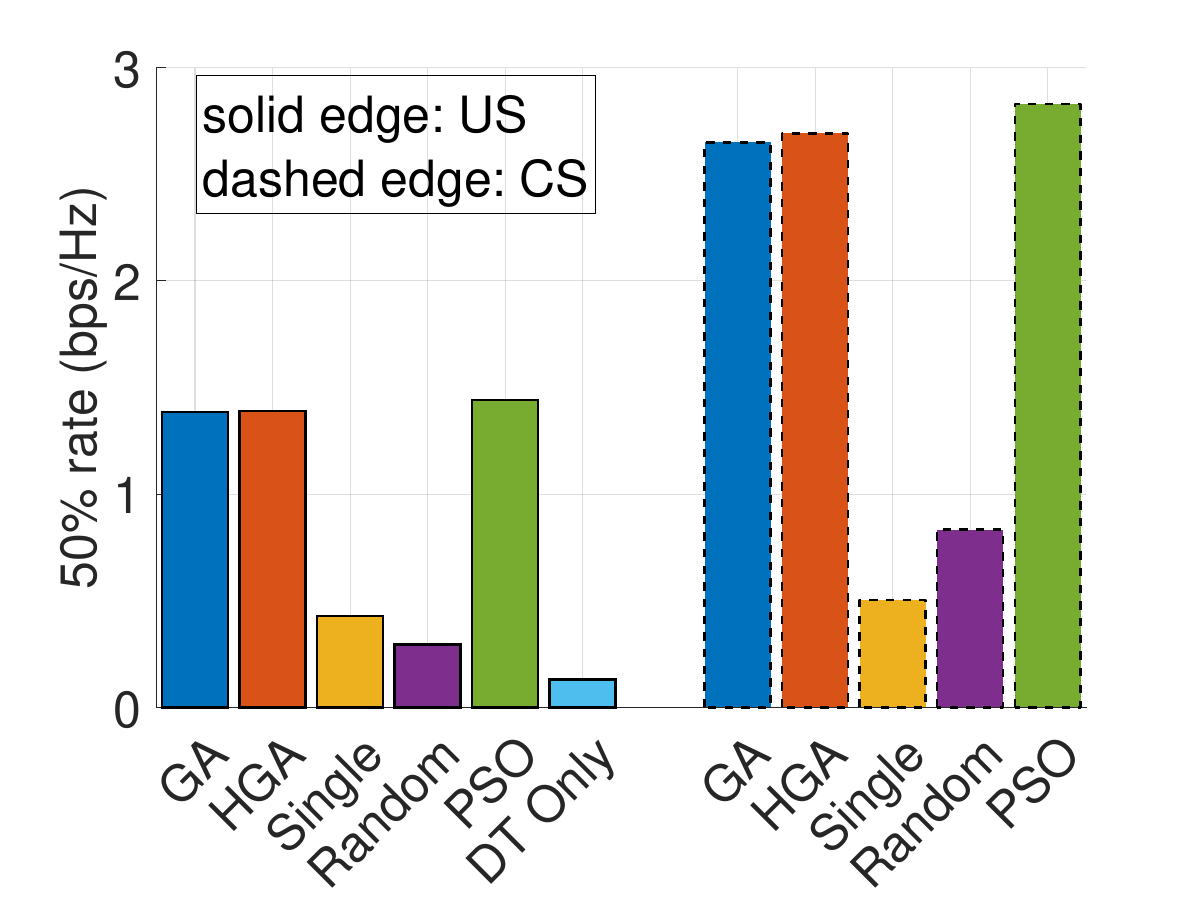}
  \caption{50\% rate}
\end{subfigure}
\begin{subfigure}[t]{0.34\textwidth}
  \centering
  \includegraphics[width=\linewidth]{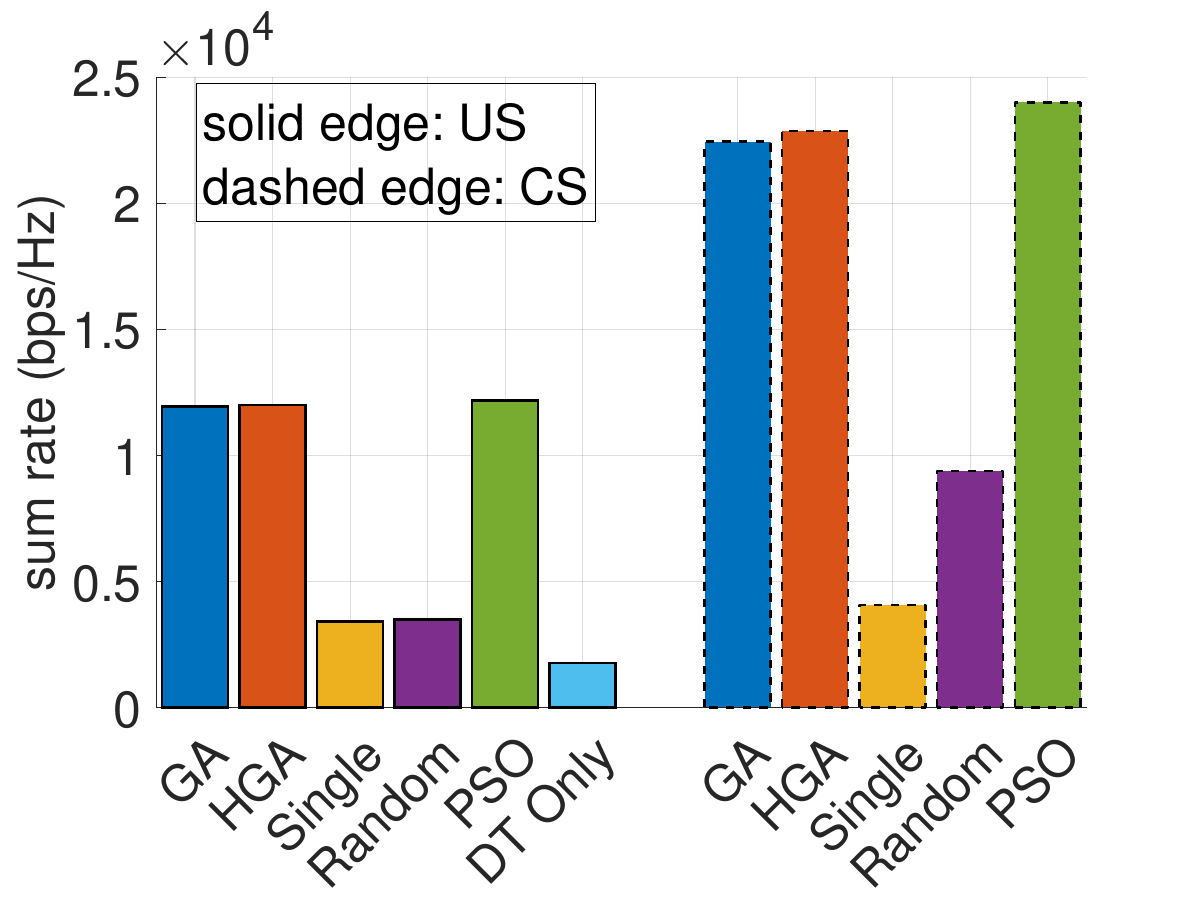}
  \caption{sum rate}
\end{subfigure}
\caption{Rate with ISD = 1000 m and $h$ = 100 m for US and CS subframes considering different algorithms.}
\label{fig:rate_1000m_100m}
\vspace{-.2in}
\end{figure*}

Fig.~\ref{fig:rate_500m_200m} and Fig.~\ref{fig:rate_1000m_100m} show the minimum rate, median rate, and sum rate performance of various uptilt optimization techniques in both US and CS scenarios. For the denser network with $\mathrm{ISD} = 500$ m shown in Fig.~\ref{fig:rate_500m_200m}, the network is in the interference-limited regime due to the presence of multiple dominant line-of-sight interferers. Hence, for the US case, where both UT and DT sectors transmit simultaneously, the random and single tilt techniques have the worst performance in all cases. On the other hand, the GA-based techniques have the best performance in terms of both minimum and median rate due to the coordination of the uptilt angles.

Compared to all the techniques, the PSO technique has the highest minimum, median, and sum rate performance. This shows that the PSO technique is able to efficiently traverse the non-convex optimization space and find well-structured tilt configurations. For the case where data transmissions are muted during the CS slots, it is evident that all the techniques benefit from the performance improvement due to the reduced DT sidelobe interference. The benefits are more noticeable in the optimized techniques.

For the case of $\mathrm{ISD} = 1000$ m in Fig.~\ref{fig:rate_1000m_100m}, the overall rate increases because of the lower inter-cell interference; however, the performance of the random and single-tilt schemes is still poor in comparison to the coordinated optimization techniques. PSO is still the best approach, while coordinated slots outperform uncoordinated slots, though by a smaller margin than in the dense deployment case.

\begin{figure*}[t]
\centering
\begin{subfigure}[t]{0.325\textwidth}
  \centering
  \includegraphics[width=\linewidth]{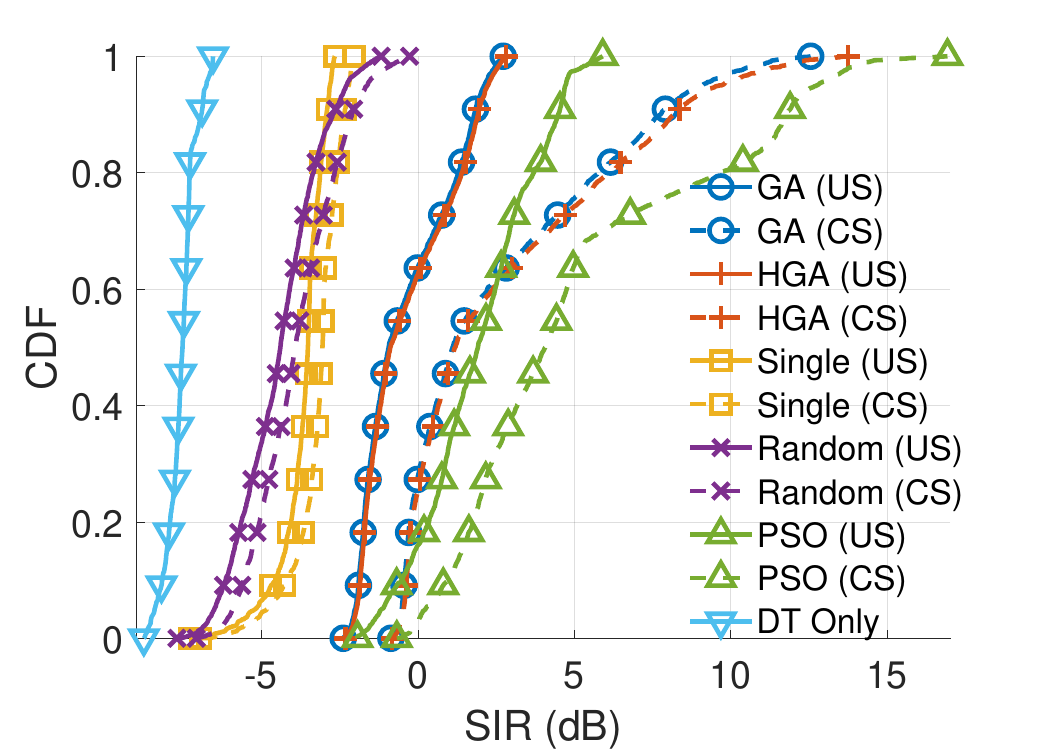}
  \caption{$h$ = 200 m, ISD = 500 m}
\end{subfigure}
\begin{subfigure}[t]{0.325\textwidth}
  \centering
  \includegraphics[width=\linewidth]{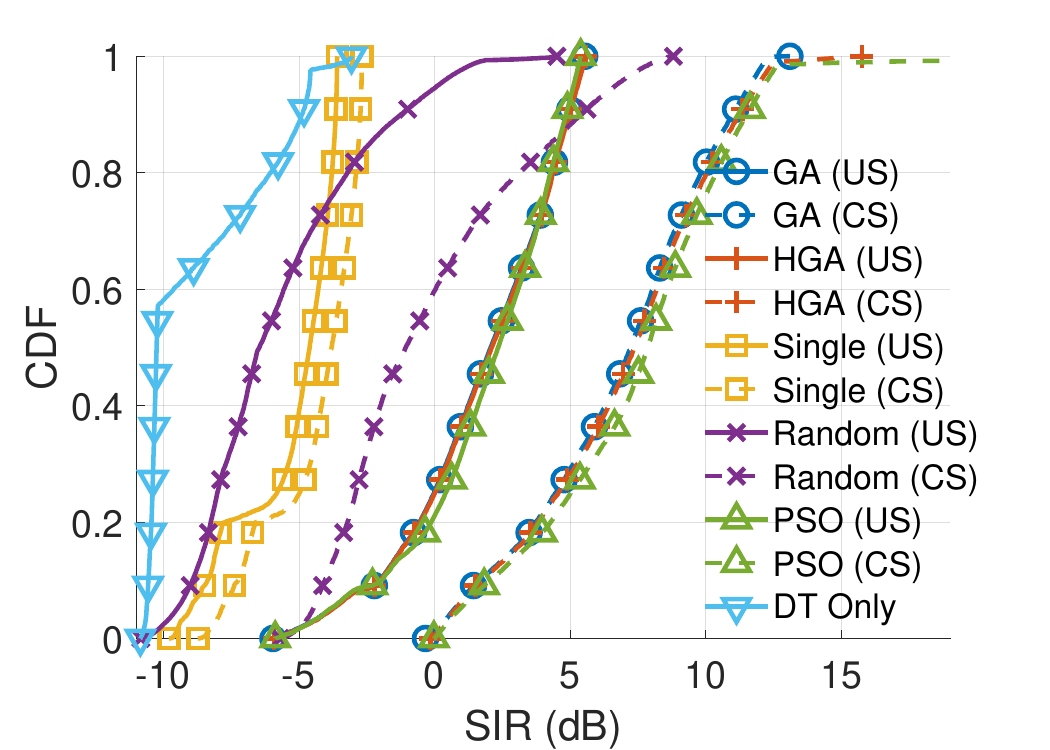}
  \caption{$h$ = 100 m, ISD = 1000 m}
\end{subfigure}
\begin{subfigure}[t]{0.325\textwidth}
  \centering
  \includegraphics[width=\linewidth]{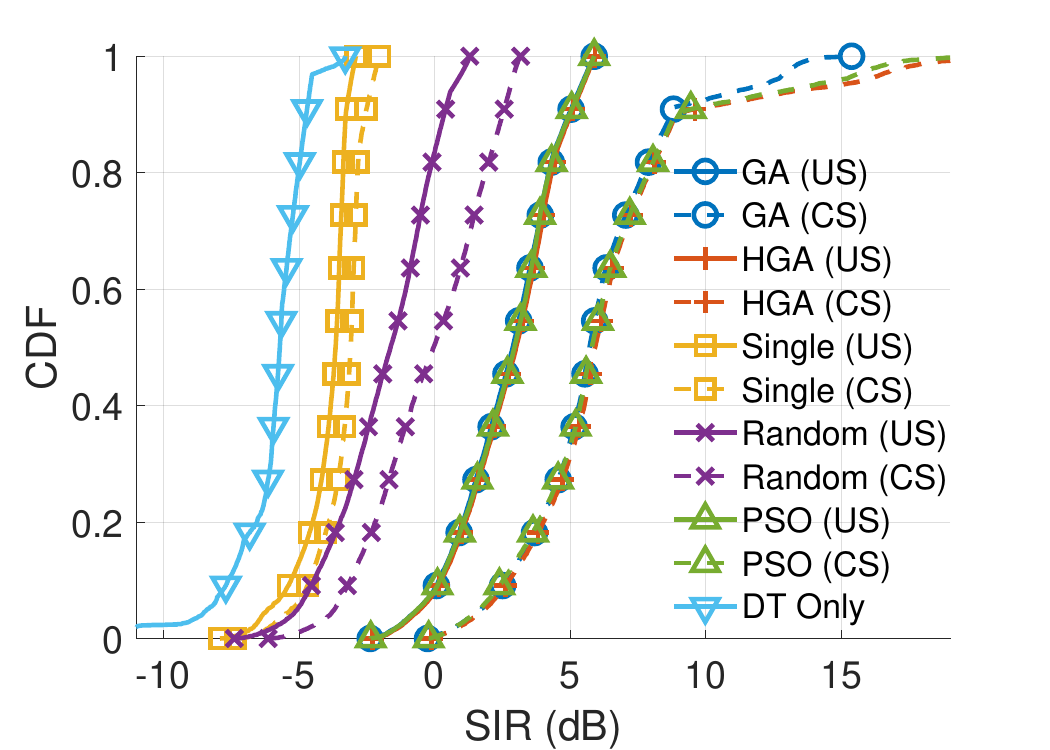}
  \caption{$h$ = 200 m, ISD = 1000 m}
\end{subfigure}
\caption{CDF of SIR with different ISDs and UAV altitudes for different optimization approaches. The solid lines represent the distribution of SIR over US resources, while the dashed lines represent the distribution of SIR over CS resources.}
\label{fig:sir_cdf_isd_500_1000}
\vspace{-.2in}
\end{figure*}

\begin{figure}[h]
\centering
\begin{subfigure}[t]{0.23\textwidth}
  \centering
  \includegraphics[width=1.9in,trim=.3in .05in 0in 0.2in,clip]{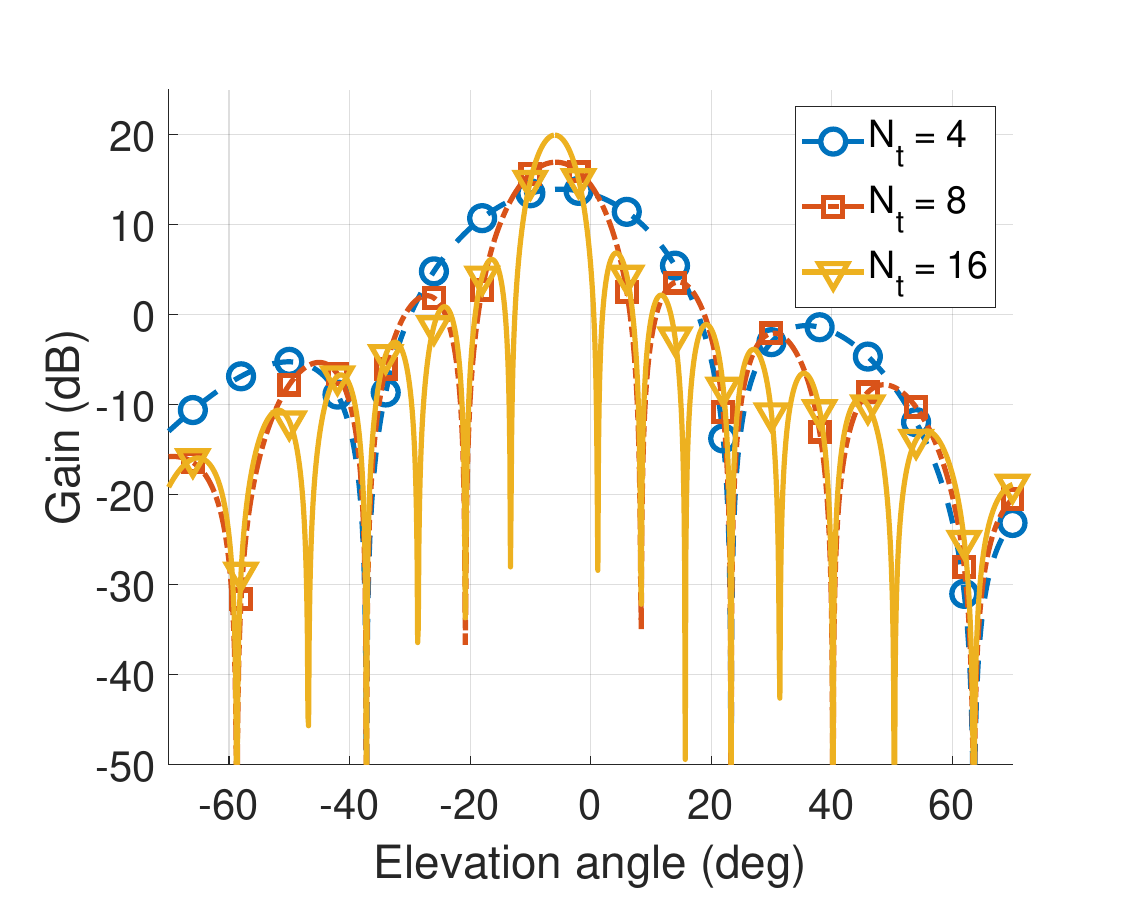}
  \caption{Vertical antenna pattern}
\end{subfigure}
\begin{subfigure}[t]{0.22\textwidth}
  \centering
  \includegraphics[width=1.8in,trim=.1in .05in 0in 0.2in,clip]{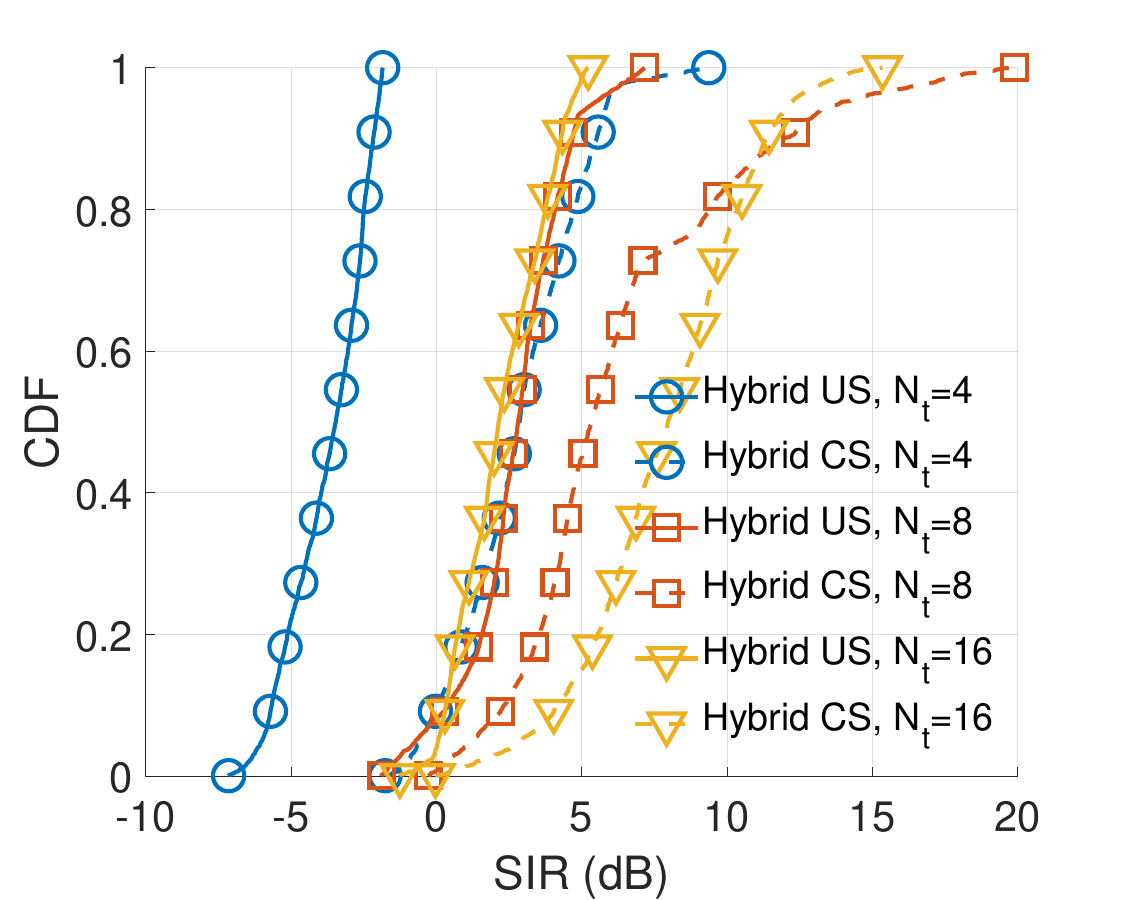}
  \caption{ISD = 500 m, $h$ = 150 m }
\end{subfigure}
\caption{Antenna gain and CDF of SIR (hybrid GA approach) for $\phi_\mathrm{dt}=-6^\circ$ with increasing $N_t$.}
\label{fig:impact_antenna_elements}
\vspace{-.2in}
\end{figure}

Fig.~\ref{fig:sir_cdf_isd_500_1000} illustrates the comparison of the SIR distribution for different uptilt optimization methods along with the DT-only baseline (where no uptilt sectors are used at any BS). For all scenarios of ISD and altitude, DT-only always experiences the worst SIR performance. This confirms that DT-only macro deployments are not adequate for UAV connectivity. Adding an uptilted sector always provides better SIR distribution, and using coordination further improves the worst-case scenario. Among all methods, PSO provides the best improvement in the minimum SIR. For dense scenarios and higher UAV altitudes, the benefit of using coordination slots is significant.

\subsection{Impact of Antenna Elements}
Fig.~\ref{fig:impact_antenna_elements} demonstrates the effect of the number of vertical antenna elements $(N_t)$ on the reliability of an aerial link. An increase in $(N_t)$ yields a greater peak gain and a narrower main lobe for the vertical antenna response. But this also contributes to sharper side lobes. This is observed for the SIR CDFs in Fig.~\ref{fig:impact_antenna_elements}(b), where an increase in $N_t$ shifts the SIR CDF to the right under both US and CS slots, indicating improved link quality over the UAV grid. In addition, by applying TDIC, the CS slots outperform the US slots for all values of $N_t$ .

\subsection{Impact of Downtilt and Duty Cycle on GUE}
\begin{figure}[t]
\centering
\begin{subfigure}[t]{0.21\textwidth}
  \centering  
  \includegraphics[width=1.8in,trim=0in 0in 0in 0in,clip]{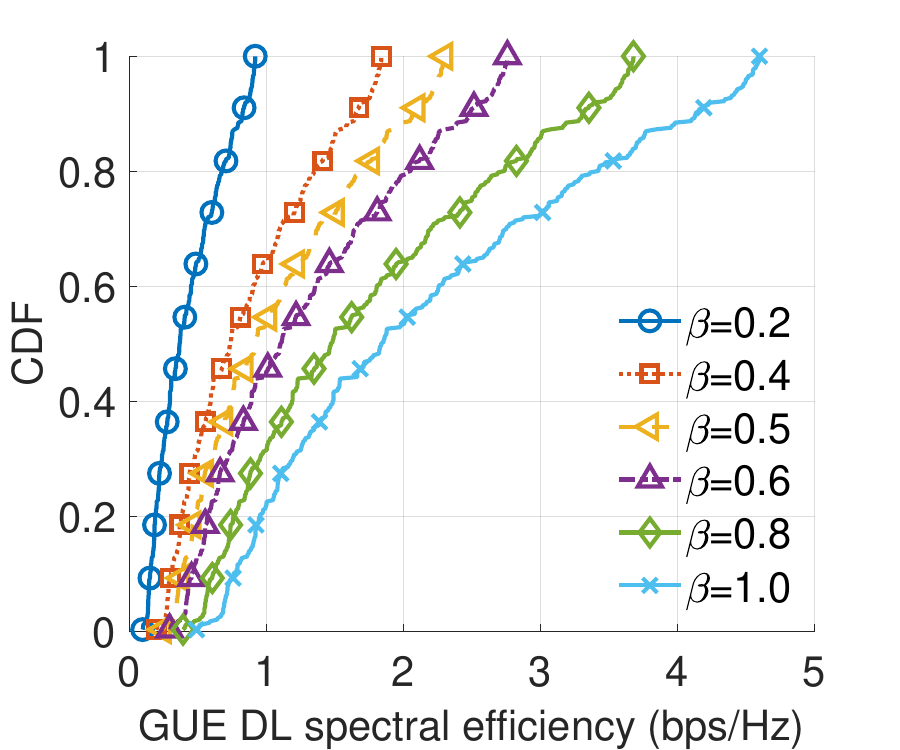}
  \caption{CDF of GUE downlink spectral efficiency for different $\beta$.}
\end{subfigure}
\hspace{.1in}
\begin{subfigure}[t]{0.23\textwidth}
  \centering
  \includegraphics[width=1.8in,trim=.1in 0.23in 0.2in 0in,clip]{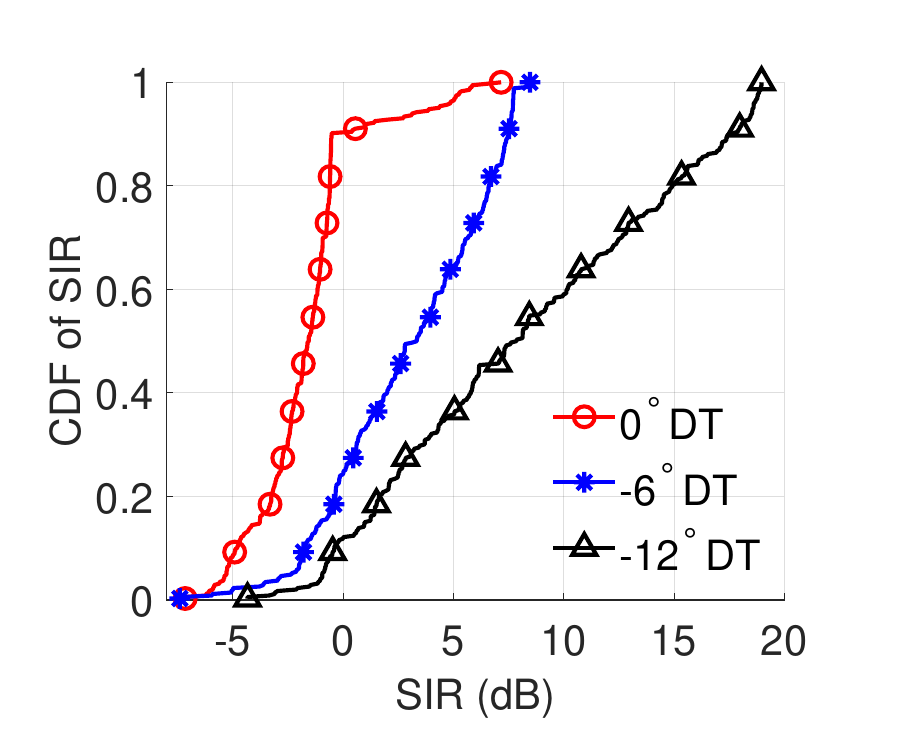}
  \caption{CDF of GUE SIR for different down-tilt angles, $h_{\mathrm{g}}=1.5$~m).}
\end{subfigure}

\caption{Impact of antenna down-tilt and  duty cycle on GUE performance under DT-active transmission for ISD = 100 m.}
\label{fig:gue_impact}
\vspace{-0.2cm}
\end{figure}

Fig.~\ref{fig:gue_impact} illustrates the impact of  inter-cell interference
coordination (ICIC) duty cycle and antenna downtilt on the performance of the GUEs when DT-active transmission is employed. Fig.~\ref{fig:gue_impact}(a) shows the CDF of the GUE's downlink spectral efficiency for different values of the ICIC duty cycles, $\beta$. We observe that as the value of $\beta$ increases, the fraction of the available time resource allocated to the DT transmission serving the GUEs increases, resulting in the right shift of the corresponding CDF. Also, from the lower tail of the CDF, it is clear that the increased value of $\beta$ not only benefits the GUE's average throughput but also the cell edge users. On the other hand, for lower values of $\beta$, the GUE's spectral efficiency decreases, reflecting the cost in GUE throughput due to the increased fraction of the available resource allocated for mitigating interference towards the UAV users.

Fig.~\ref{fig:gue_impact}(b) shows the CDF of the GUE SIR for various antenna downtilt values, $\phi_\mathrm{dt}$. Increasing the magnitude of $\phi_\mathrm{dt}$ from $0^\circ$ to $-12^\circ$ shifts the CDF of the GUE SIR to the right, showing an improvement in SIR values. We can see that lower values of $\phi_\mathrm{dt}$ (antenna more tilted towards the ground) significantly improve the distribution of the GUE SIR by better aligning the main antenna lobe with the ground users and minimizing interference between cells. The constant rightward shift of the CDF for increasing down-tilt values verifies the accuracy of the proposed elevation angle geometry and antenna pattern.

\section{Conclusion and Future Works}
\label{sec:conclusion}
In this work, we present a deployable, network-side approach to improve the downlink reliability of cellular-connected UAVs in multi-cell 5G NR macro networks. We combine BS uptilt optimization with NR-compatible time-domain interference coordination (TDIC) without requiring any changes to the 5G NR air interface. By focusing on a worst-case SIR objective, the proposed uptilt optimization approaches directly address the interference-limited nature of aerial downlink links. From simulations based on a 19-cell wraparound model, we observe that an additional booster cell with TDIC and an uptilt optimization consistently outperforms downtilt-only sector, single-tilt, and random-tilt baselines in terms of worst-case SIR and downlink sum rate. The hybrid GA provides more reliable performance than the standard GA. PSO achieves the best worst-case SIR across all evaluated scenarios. In addition, the TDIC approach further improves performance across all considered scenarios. As future work, we plan to incorporate dynamic UAV mobility, imperfect channel state information, and near-real-time RAN intelligent controller (RIC)-based control to enable closed-loop aerial optimization in 5G and 6G networks. We also plan to adaptively select the duty-cycle parameter, $\beta$, based on traffic load and network conditions to ensure that ground users' QoS is maintained regardless of UAV traffic demands.
\bibliographystyle{IEEEtran}
\bibliography{references}
\end{document}